# Preliminary Analysis of Channel Capacity in Air-to-ground LoS MIMO Communication Based on A Cloud Modeling Method

Ning Wei, Shuangqing Tang, Zeyuan Zhang

University of Electronic Science and Technology of China, Oct. 2022

## ABSTRACT

Since the orthogonality of the line-of-sight multiple input multiple output (LoS MIMO) channel is only available within the Rayleigh distance, coverage of communication systems is restricted due to the finite implementation spacing of antennas. However, media with different permittivity in the transmission path are likely to loosen the requirement for antenna spacing. Such a conclusion could be enlightening in an air-to-ground LoS MIMO scenario considering the existence of clouds in the troposphere. To analyze the random phase variations in the presence of a single-layer cloud, we propose and modify a new cloud modeling method fit for LoS MIMO scene based on real-measurement data. Then, the preliminary analysis of channel capacity is conducted based on the simulation result.

## 1 Introduction

With the increasing range of activity of mankind, long-range networking over the air is one of the prevailing trends in next-generation communications. Under this circumstance, LoS MIMO is a commonly considered system model to reach high spectrum efficiency. Though the antenna arrangement that yields ideal spatial multiplexing can be determined for a variety of array geometries [1-4], the orthogonality of LoS MIMO systems is restricted within the Rayleigh distance [5]. On the other hand, meteorological events are one of the major concerns in outdoor communications, especially in LoS MIMO systems that are commonly considered in air-to-ground communication links. ITU has proposed a calculation model for designers to estimate the antennation due to clouds [6]. Nevertheless, few works have considered cloud modeling and phase variation caused by clouds under the LoS MIMO scene. In addition, as mentioned in [7], to further extend the communication range of LoS MIMO systems, it is also noteworthy to explore channel characteristics beyond the Rayleigh distance.

To gain a deeper knowledge of the influence introduced by monolayer cloud on the air-to-ground link, i.e., the most general scene. We propose a spherical cloudlets model on the basis of E-band real-measured data and further conduct performance analysis of channel capacity both for transmission distance within and beyond the Rayleigh distance.

## 2 Cloud Modeling

We consider the cloud layer as a rectangular media region with inhomogeneous distribution of ice and water content (IWC). As shown in Figure 1. Assume the width and the height of the rectangular region are $W$ and $D$, respectively.

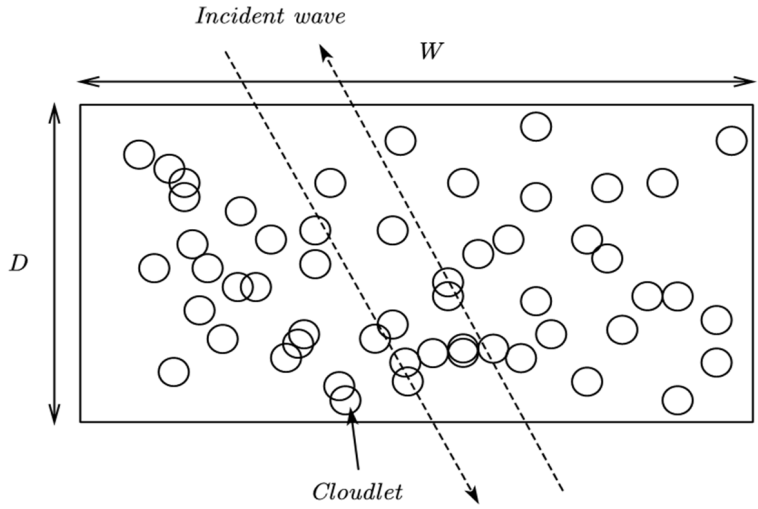

Figure 1. Schematic diagram of the spherical cloudlets model

In the rectangular region, inclusions of spherical cloudlets that manifest the inhomogeneous distribution of ice and water content are performed based on the Poisson point process with Poisson density $\lambda_s WD$. We assume that the radius of each cloudlet is identical, denoted as

$$r = \frac{\alpha W \sqrt{\frac{D}{D_{max}}}}{2} \qquad (1)$$

Where parameter $\alpha$ is introduced as the smoothness factor to represent the correlation between spherical cloudlets, which manifests the dependence of IWC. $D_{max}$ represents the maximum cloud thickness. The IWC $C_c$ in cloudlet is considered uniformly distributed, i.e., $C_c \sim U(0, C)$, where $C$ denotes the maximum IWC available in a single layer cloud. In this model, the IWC at any location within the rectangular region is determined jointly by the spherical cloudlets that produce overlap at that location, which is in accordance with the practical situation of cloudlet distribution.

The time-varying effect of the IWC is applied to the model by endowing velocity $V_b$ to each cloudlet, we consider the location variation $\Delta x_{\Delta t}$ and $\Delta y_{\Delta t}$ during time $\Delta t$ follows uniform distribution $U(-V_b \Delta t, V_b \Delta t)$. The location variation is programmed to take inverse if the movement in $\Delta t$ exceeds the rectangular region.

To improve the effectiveness of the proposed spherical cloudlets model, we present function fitting to the model according to real-measurement data. The data, i.e., three groups of polarization-dependent loss (PDL) value, each with 3000 samples, were acquired in a 20 km air-to-ground dual polarized transmission link operated at 73.5 GHz. Figure 2 shows the cumulative distribution function (CDF) of PDL of measurement result and the fitted spherical cloudlets model. All parameter used in the simulation is shown in Table 1.

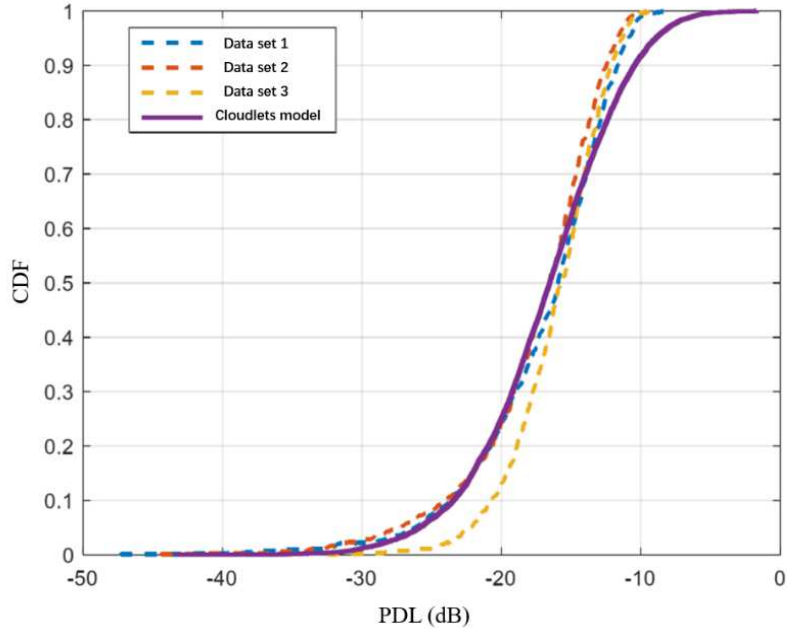

Figure 2. PDL CDF of the measurement data and proposed model

Table 1. Parameter setup of the simulation

| Item | Value |
|---|---|
| Poisson density $\lambda_s$ | 0.002 |
| Density of cloudlets | 30000 $pcs/m^3$ |
| IWC | 0.4 $g/m^3$ |
| Cloudlets velocity $V_b$ | 1000 $m/s$ |
| Elevation of the measurement link | 85.14° |
| Thickness of the cloud layer | 1000 $m$ |

In comparison to the existing cloud modeling method in the literature, e.g., cloud sense simulation model (CSSM) in [8] and [9], the spherical cloudlets model has a huge reduction in complexity. For example, when resolution of the CSSM is set as 10 m, the average multiply accumulate (MAC) operation over 1000 times simulation of the CSSM and cloudlets model are shown in Table 2. This is mainly because the CSSM requires fractal interpolation operation to generate multiple physical parameters of clouds and requires coverage more than hundreds of square meters in single generation. While the spherical cloudlets model eliminates the complexity of interpolation and enables precise concentration on a small coverage of regard.

Table 2. MAC comparison for different models

| Modeling method | Cloud thickness | Coverage | MAC per round |
|---|---|---|---|
| CSSM | 1000 $m$ | 500 $m^2$ | 5391772 |
| Proposed model | 1000 $m$ | 20 $m^2$ | 10367 |

Furthermore, the overlapping of cloudlets in spherical cloudlets model enables continuous phase variation in comparison to the grid-based cloud modeling (see Figure 3), which is more suitable for communication applications.

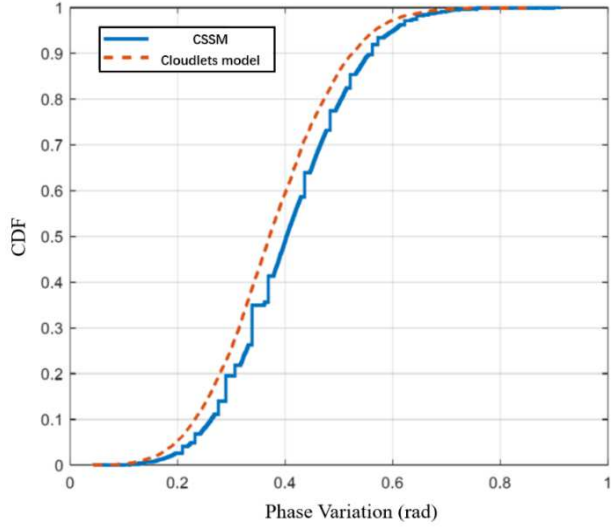

Figure 3. Additional phase variation CDF of the CSMM and proposed model

3 Analysis of The Phase Variation

In this part, we further verify the effectiveness of the proposed spherical cloudlets model by conducting phase variation analysis. Note that phase variation introduced by the cloud can be represented as

$$\phi_c = \sum_{k=0}^{\infty} P_k \phi_{c_k} \tag{2}$$

Where $\phi_{c_k}$ denotes the total phase variation introduced by k cloudlets, while $P_k$ denotes the probability of k involved cloudlets. The distribution of $\phi_c$, is considered Gaussian-like, according to Liapunov central limit theorem (CLT). Yet considering the phase variation ranges from 0 to $2\pi$, the distribution is characterized by leptokurtosis and fat-tail. Thus, we use Laplace distribution as an approximate distribution of $\phi_c$, i.e.,

$$f(\phi_c) = \frac{1}{\sqrt{2\sigma_c}} e^{\left[-\frac{|\phi_c - \phi_0|}{\sqrt{\sigma_c/2}}\right]} \tag{3}$$

Where

$$\phi_0 = \sum_{k=0}^{\infty} P_k E(\phi_{c_k}) \tag{4}$$

$$\sigma_c^2 = \sum_{k=0}^{\infty} P_k^2 D(\phi_{c_k}) \tag{5}$$

To calculate $E(\phi_{c_k})$ and $D(\phi_{c_k})$, the complex permittivity with respect to free space of the ice cloud $\tilde{\varepsilon}_{ice}$ can be obtained by considering the Rayleigh mixture [10]:

$$\tilde{\varepsilon}_{ice} = 3n V_{ice} \frac{C_b}{0.6} \frac{\varepsilon_{ice}' - 1}{\varepsilon_{ice}' + 1} \tag{6}$$

Where $n$ denotes the density of the medium sphere, $V_{ice}$ denotes the volume of the ice sphere, $\varepsilon_{ice}'$ is the real part of the complex permittivity of the ice sphere. According to [11], the wavelength in ice sphere can be calculated as

$$\lambda = \frac{\lambda_0}{1+\tilde{\varepsilon}_{ice}} \qquad (7)$$

With $\lambda_0$ denotes the wavelength in free space, the additional phase variation obtained from transmission distance $l$ in the medium sphere, i.e.,

$$\varphi_{add} = \left(\frac{2\pi}{\lambda} - \frac{2\pi}{\lambda_0}\right) \cdot l = \frac{2\pi l}{\lambda_0}\tilde{\varepsilon}_{ice} \qquad (8)$$

We can now derive $E(\phi_{c_k})$ and $D(\phi_{c_k})$ as

$$E(\phi_{c_k}) = E\left(\sum_{i=1}^{k}\frac{2\pi l_i}{\lambda_0}\varepsilon_i\right) = \frac{2\pi}{\lambda_0}\sum_{i=1}^{k}E(\varepsilon_i l_i) = \frac{2\pi k}{\lambda_0}E(l)E(\varepsilon) \qquad (9)$$

$$D(\phi_{c_k}) = \left(\frac{2\pi}{\lambda_0}\right)^2 k^2(E(\varepsilon^2)E(l^2) - E^2(\varepsilon)E^2(l)) \qquad (10)$$

Where $\varepsilon$ and $l$ denote the relative permittivity and the length involved in the traveling path of the cloudlets, while $\varepsilon_i$ and $l_i$ denote the relative permittivity and the length involved in the traveling path of the i-th cloudlet. With $E(\varepsilon)$ and $E(\varepsilon^2)$ obtained from equation (6) as

$$E(\varepsilon) = 3nV_{ice}\frac{1}{0.6}\frac{\varepsilon_{ice}'-1}{\varepsilon_{ice}'+1}E(C_c) \qquad (11)$$

$$E(\varepsilon^2) = \left(3nV_{ice}\frac{1}{0.6}\frac{\varepsilon_{ice}'-1}{\varepsilon_{ice}'+1}\right)^2 E(C_c^2) \qquad (12)$$

With respect to $l$, the probability density function can be approximated by direct curve fitting, the raw moment can be written as

$$E(l) = \frac{L}{k_l^2}[e^{2k_l r}(2k_l r - 1) + 1] \qquad (13)$$

$$E(l^2) = \frac{L}{k_l^3}[e^{2k_l r}(4k_l^2 r^2 - 4rk_l^3 + 2) - 2] \qquad (14)$$

Where $k_l = 2.35 \times 0.7^{\log_2 \frac{D}{D_{max}}}$. We further approximate $P_k$ according to the measurement result with function

$$P_k = 0.12 \times 2^{\log_2 \frac{D}{D_{max}}} \times e^{-\frac{\left(-k-\frac{\lambda_s \frac{D}{D_{max}}}{4\times 3 \log_2 \frac{D}{D_{max}}}\right)^2}{30}} \qquad (15)$$

Now substitute (9)-(15) into (4) and (5), the distribution of stationary phase variation $\phi_c$ can be obtained. The time-varying component $\Delta t$ can be further considered on the basis of stationary phase variation model, i.e., $\phi_{total} = \phi_c + \phi_{\Delta t}$. Since the derivation process of phase perturbation $\phi_{\Delta t}$ is similar to that of stationary phase variation, we directly give distribution of $\phi_{total}$ without redundant analysis (see Appendix A for full expression).

## 4 Discussion on Channel Capacity

In air-to-ground LoS MIMO systems, the channel capacity depends strongly on the orthogonality of the sub-channel, which can be obtained by the proper arrangement of the transceiver antennas. However, the orthogonality is fortuitous due to the variation in the link path. With the application of spherical cloudlets model, we can now focus on the channel performance in the presence of monolayer cloud. We set the simulation parameters as shown in Table 3.

Table 3. Parameter setup for LoS MIMO system

| Parameter | Value | Parameter | Value |
|---|---|---|---|
| Number of transmit antenna | 2 | Operation frequency | 73.5 GHz |
| Number of receive antenna | 2 | Spacing between transmit antenna | 1 m |
| Average signal-to-noise ratio (SNR) | 20 dB | Spacing between transmit antenna | 6.0827 m |
| Transmission distance | 10 km/40 km | Upper bound of cloud layer | 8 km |

The 10 km and 40 km are set as benchmarks of transmission distance within and beyond the Rayleigh distance. Figure and show the CDF of channel capacity concerning the relative water content (RWC), i.e., the ratio between water content and maximum water content ($0.6 g/m^3$).

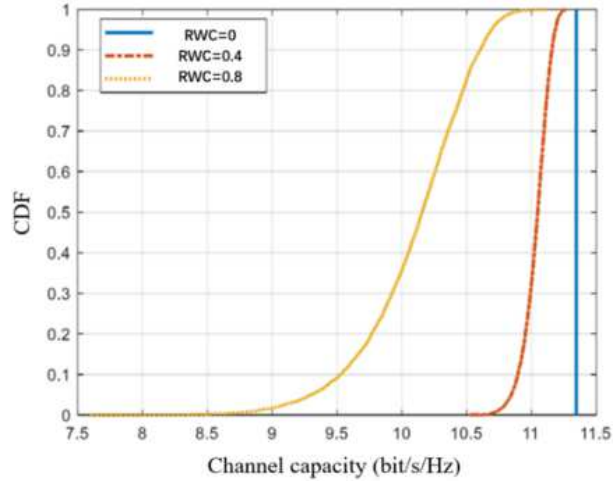

Figure 4. CDF of channel capacity with different RWC at 10 km

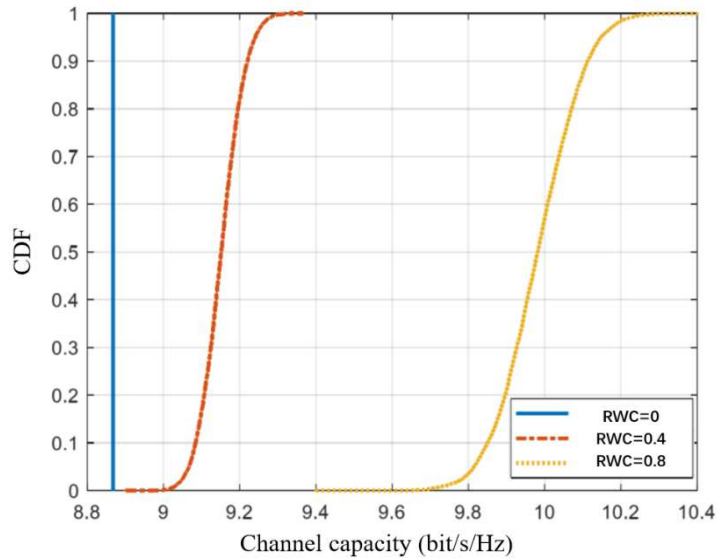

Figure 5. CDF of channel capacity with different RWC at 40 km

As can be observed in Figure 4, when transmission distance is 10 km, the outage capacity under 0.5 probability of the RWC value 0, 0.4 and 0.8 is 1.4 bit/s/Hz, 11.1 bit/s/Hz, and 10.2 bit/s/Hz, respectively. Since greater the RWC value result in greater the variance of the phase variation of each channel, and thus the greater the impact on the rank of the channel matrix. We can conclude that RWC draws an adverse impact on the channel capacity within the Rayleigh distance.

While as illustrated in Figure 5, when transmission distance is 40 km, the outage capacity under 0.5 probability of the RWC value 0, 0.4 and 0.8 is 8.88 bit/s/Hz, 9.15 bit/s/Hz, and 9.95 bit/s/Hz, respectively. Counterintuitively, the increase in RWC tends to improve the system performance beyond the Rayleigh distance. Similar conclusion fit with the thickness of the cloud $D$. Figure 6 and 7 show the CDF of channel capacity with respect to different $D$. The thickness of the cloud impairs the channel capacity within the Rayleigh distance while can improve the channel capacity beyond the Rayleigh distance.

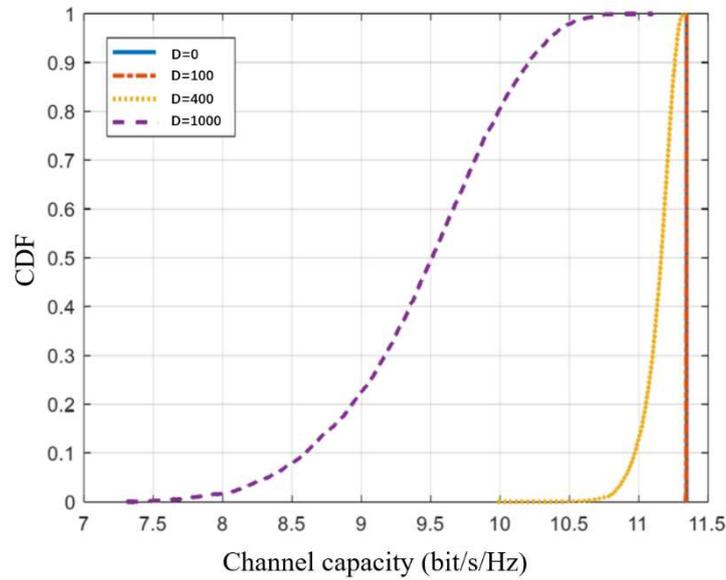

Figure 6. CDF of channel capacity with different D at 10 km

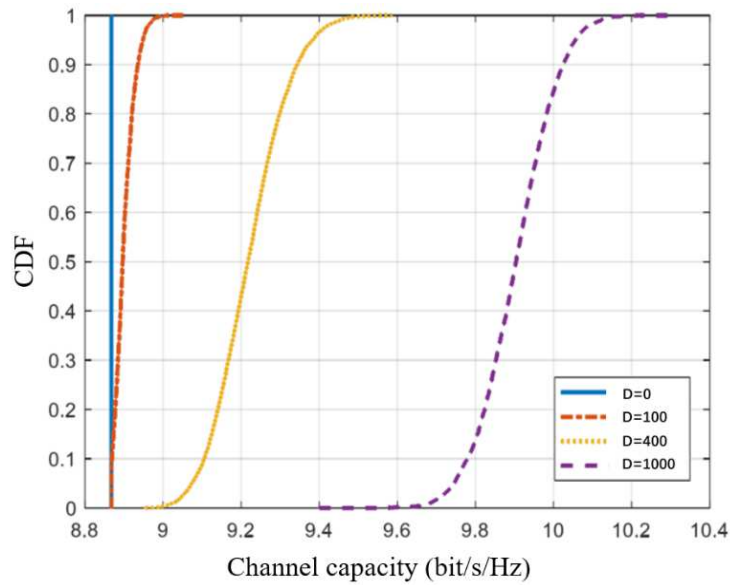

Figure 7. CDF of channel capacity with different D at 40 km

We further analyze the correlation between sub-channels in the presence of monolayer cloud, configuration of the LoS MIMO is shown in Table 4, while the parameter of the cloudlets model is same as Table 1.

Table 4. Parameter setup of the LoS MIMO system

| Parameter | Value | Parameter | Value |
|---|---|---|---|
| Number of transmit antenna | 2 | Operation frequency | 73.5 GHz |
| Number of receive antenna | 2 | Spacing between transmit antenna | 1 m |
| Average signal-to-noise ratio (SNR) | 20 dB | Spacing between transmit antenna | 5 m |
| Transmission distance | 0-30 km | Upper bound of cloud layer | 8 km |

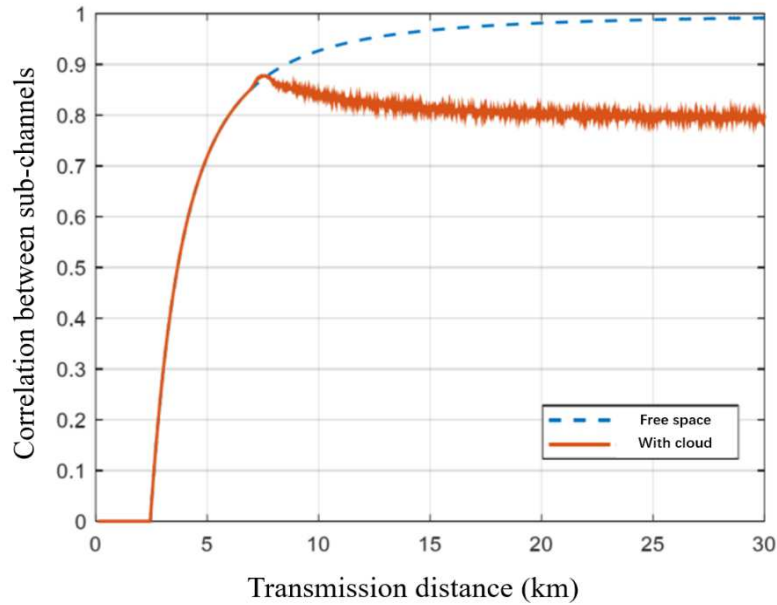

Figure 8. Correlation between sub-channels w/ and w/o presence of cloud

Figure 8 shows the correlation between sub-channels is reduced due to the presence of cloud layer when reached the height of lower bound. If we consider an energy-sufficient scenario where the loss due to the increase of transmission distance can be compensated, that is, we consider identical received power at different distance. We can then find the channel capacity as shown in Figure 9.

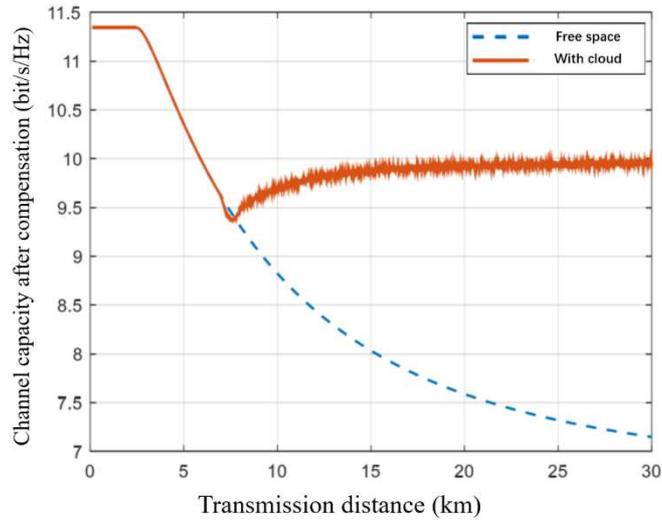

Figure 9. Channel capacity after compensation

This means that when the transmission distance excesses the Rayleigh distance, the presence of monolayer cloud be considered a make-up to the channel capacity deterioration.

## 5 Conclusion

In this paper, firstly, we have proposed a spherical cloudlets model in air-to-ground LoS MIMO scene on the basis of real-measured data, which suit for communication analysis. The complexity is reduced largely compared to the meteorologic method that exists in the literature. We then analyze the performance of channel capacity in the presence of a monolayer cloud both within and beyond the Rayleigh distance. The outcome shows cloud in the transmission link can be utilized to compensated for loss in capacity. Optimization of the proposed model maybe further extended by the study of in-depth statistical characteristic of the cloud parameters, more importantly, specific method to make use of the existing medium in the transmission link to extend communication range is a noteworthy topic.

Appendix A. Full distribution expression of the phase vitiation introduced by monolayer cloud.

$$\phi_{total} = \phi_c + \phi_{\Delta t}$$

$$f(\phi_c) = \frac{1}{\sqrt{2\sigma_c}} e^{\left[-\frac{|\phi_c - \phi_0|}{\sqrt{\sigma_c/2}}\right]}$$

Where

$$\phi_0 = \sum_{k=0}^{\infty} P_k E(\phi_{c_k}) = 0.12 \times 2^{\log_2 \frac{D}{D_{max}}} \times \frac{2\pi}{\lambda_0} E(l) E(\varepsilon) \times \sum_{k=0}^{\infty} k e^{-\left(\frac{k-B_2}{5.5}\right)^2}$$

$$= \frac{1500 n \pi V_{ice}(\varepsilon_{ice} - 1)}{\lambda_0(\varepsilon_{ice} + 1)} \times C \times B_1 e^{-\left(\frac{B_2 - 34}{14}\right)^2} \times \left(\frac{L\left((2k_l - 1) \times e^{k_l} + 1\right)}{k_l^2}\right)$$

$$\sigma_c^2 = \sum_{k=0}^{\infty} P_k^2 D(\phi_{c_k}) = \left(\frac{2\pi B_1}{\lambda_0}\right)^2 (E(\varepsilon^2)E(l^2) - E^2(\varepsilon)E^2(l)) \sum_{k=0}^{\infty} k^2 e^{-2\left(\frac{k-B_2}{5.5}\right)^2}$$

$$= \left(\frac{98 n \pi V_{ice}(\varepsilon_{ice} - 1) C B_1}{\lambda_0(\varepsilon_{ice} + 1)}\right)^2 \times \frac{L}{k_l^3} e^{-\left(\frac{B_2 - 36.7}{12}\right)^2}$$

$$\times \left(16.6 e^{2k_l r}(2k_l^2 r^2 - 2k_l^3 r - 1.5lr)\right.$$

$$\left. + 6.25L \times e^{4k_l r}(-4k_l r^2 + 4r - k_l^{-1}) - 16.6 \left(\frac{6.25L + k_l}{k_l}\right)\right)$$

Where $L = 5 \times 10^{-16}$, $B_1 = 0.12 \times 2^{\log_2 \frac{D}{D_{max}}}$, $B_2 = \left(\frac{\lambda_s D}{D_{max}} \times 3^{\log_2 \frac{D}{D_{max}}}\right)/4$

$$f(\phi_{\Delta t}) = \frac{1}{\sqrt{2\pi \Delta \sigma_c^2}} e^{\left(-\frac{(\phi_{\Delta t} - \Delta \phi_0)^2}{2 \Delta \sigma_c^2}\right)}$$

Where $\Delta \phi_0 = 0$

$$\Delta \sigma_c^2 = \sum_{k=0}^{\infty} P_k^2 D(\phi_{c_k})$$

$$= 2400 e^{-\left(\frac{B_2 - 36.7}{12}\right)^2} \times \left(\frac{2\pi B_1}{\lambda_0}\right)^2 (E(\varepsilon^2)E(\Delta l^2) - E^2(\varepsilon)E^2(\Delta l))$$

$$= 2400 \sqrt{\Delta t V_b} \left(\frac{3\sqrt{3} n \pi V_{ice}(\varepsilon_{ice} - 1) C B_1}{10 \lambda_0(\varepsilon_{ice} + 1)}\right)$$

$$\times 0.9^{\log_2 \frac{D}{D_{max}}} \times e^{-\left(\frac{B_2 - 36.7}{12}\right)^2}$$

Where $\Delta l$ demotes transmission distance variation in single cloudlet within time $\Delta t$. $\Delta l$ can be obtained by curve fitting which yields:

$$f(\Delta l) = \frac{1}{\sqrt{2\pi \Delta \sigma_{\Delta l}{}^2}} e^{\left(-\frac{(\Delta l - \Delta l_0)^2}{2\Delta \sigma_{\Delta l}{}^2}\right)}$$

$$\Delta l_0 = 0$$

$$\Delta \sigma_{\Delta l}{}^2 = \sqrt{\Delta t V_b} \times 0.9^{log_2 \frac{D}{D_{max}}}$$